\documentclass[preprint,nofootinbib,aps,superscriptaddress,showpacs]{revtex4}
\usepackage{epsfig}
\usepackage{amsmath}
\usepackage{feynmf}
\usepackage{changes}
\def\be{\begin{equation}} 
\def\ee{\end{equation}} 
\def\bea{\begin{eqnarray}}
\def\eea{\end{eqnarray}}

\newcommand{\no}{\nonumber \\}

\newcommand{\del}{\partial}

\begin{document}
\title{
Spin polarization observables of the deuteron photodisintegration 
at low energies in pionless effective field theory}
\author{Young-Ho Song}
\affiliation{Rare Isotope Science Project, 
Institute for Basic Science, Daejeon 34047, Korea}
\author{Shung-Ichi Ando}
\affiliation{School of Mechanical and ICT Convergence Engineering, 
Sunmoon University, Asan, Chungnam 31460, Korea}
\author{Chang Ho Hyun}
\email{hch@daegu.ac.kr}
\affiliation{Department of Physics Education, Daegu University,
Gyeongsan 38453, Korea}

\date{June 22, 2017}

\begin{abstract}
Spin polarization observables of the deuteron 
photodisintegration at low energies
are studied in a pionless effective field theory 
up to next-to-next-to-leading order (NNLO). 
The total and differential cross sections, 
induced neutron polarization $P_{y'}$, and tensor analyzing powers
$T_{20}$ and $T_{22}$ of the process are calculated at photon energies
from the breakup threshold to 20~MeV.
We find that the NNLO corrections in the cross sections and $P_{y'}$ 
converge well whereas they turn out to be important contributions 
in $T_{20}$ and $T_{22}$.  
We discuss the discrepancy between theory and experiment
in $P_{y'}$ still persisting as well as an implication of our result 
to the first measurement of $T_{20}$ at low energies 
in the HIGS facility.

\end{abstract}


\maketitle

\section{Introduction}

Observables related with polarization provide
more detailed information on nuclear reactions  
than the unpolarized ones~\cite{Arenhovel:1990yg}. 
There still remain couples of problems 
which show discrepancy between experiment and theory 
even in few-nucleon systems 
at low energies~\cite{Schiavilla:2005zt,Kukulin:2008zz,Ando:2011nv}. 
Induced neutron polarization $P_{y'}$ of the photodisintegration of 
deuteron is an example and 
one can find more information in the HIGS2 proposal~\cite{higs}.
Proposal for the photodisintegration of deuteron 
at HIGS is focusing on the role of the final state.
Tensor analyzing power $T_{20}$ is chosen for the investigation, 
which depends on the $d$-state
of the deuteron wavefunction as well as the polarization states.
Main objective of the experiment is to figure out 
or resolve the discrepancy between experiment
and theory at photon energies around 10~MeV, 
and obtain precise data for $T_{20}$ 
which has not been measured yet in the low-energy regime.   

Pionless effective field theory (EFT) for low energy phenomena,
in which the pion can be treated as a heavy degree of
freedom and integrated out of the effective Lagrangian,
provides us a model-independent and perturbative 
calculation method~\cite{Bedaque:2002mn,Chen:1999tn,Ando:2004mm}.
An effective Lagrangian of the pionless EFT can be constructed 
by using only the symmetry property
of the system and momentum expansion in the low energy.
The pionless EFT is able to successfully explain many low energy 
properties of nuclear two-body systems with and without external 
electromagnetic probes~\cite{Chen:1999tn,Ando:2004mm,Rupak:1999rk,Ando:2005cz}.
Because the measured induced neutron polarization $P_{y'}$ 
in $\gamma+d\to {\vec n}+p$
and theoretical calculation using phenomenological 
potential model show a discrepancy~\cite{Schiavilla:2005zt,Ando:2011nv}, 
it is desirable to have a model independent calculation. 

In this work, 
we compute the photodisintegration cross section of deuteron 
by using a pionless EFT 
in dibaryon formalism~\cite{Ando:2004mm,Kaplan:1996nv,Beane:2000fi} 
up to next-to-next-to-leading order (NNLO) 
which includes $sd$ wave mixing effects 
of the deuteron wavefunction. 
Dibaryon fields which have the quantum numbers of two-nucleon systems
in either scattering or bound states are introduced to faciliate the resummation of effective
range effects to infinite order.
The details of formalism up to next-to-leading order (NLO) 
for the photodisintegration of deuteron have been reported 
in \cite{Ando:2004mm,Ando:2011nv,Song:2011yw}. 
The result of pionless EFT up to NLO 
in \cite{Ando:2011nv} shows discrepancy 
with the experimental data of $P_{y'}$~\cite{PhysRev.139.B71},
similar to the phenomenological nuclear force model calculations 
of \cite{Schiavilla:2005zt, Arenhovel:1990yg}, 
but the result is different from the 
potential model calculation with increasing photon energy. 
Even though the unpolarized cross sections could be well described up to NLO, 
the spin-dependent observables can be sensitive to the 
higher order corrections. 
The $d$ state in the deuteron wavefunction 
could be properly accounted when we increase the expansion
up to NNLO.
An NNLO calculation in the pionless EFT with dibaryon formalism
for the electrodisintegration of deuteron has been reported
by Christlmeier and Grie\ss hammer~\cite{Christlmeier:2008ye}. 
Because the photodisintegration of deuteron has the same electromagnetic 
hadronic currents of the electrodisintegration of deuteron up to NNLO, 
we employ the expression of the hadronic currents 
reported in \cite{Christlmeier:2008ye}.
Thus, we calculate the total cross section, the differential cross section, 
$P_{y'}$, $T_{20}$ and $T_{22}$ from threshold to 19.8 MeV in photon energy
in the center of mass frame. 
We compare the results up to NNLO with those of NLO, 
and other theoretical results. 
Experimental data are compared when available.

The paper is organized as follows.
In Sect.~\ref{sec2} we present basic formalism of the pionless EFT 
and the analytic forms of the transition amplitudes and probabilities.
In Sect.~\ref{sec3} numerical results and related discussions are given.
In Sect.~\ref{sec4} we summarize the present work.

\section{Formalism}
\label{sec2}

In this work, we employ the standard counting rules 
of the pionless EFT, whose expansion parameter,
typical momentum divided by the pion mass, satisfies
$Q\sim 1/3$~\cite{Bedaque:2002mn,Chen:1999tn}.
In this section, we display the Lagrangian up to NNLO and 
the expression of the cross sections, and obtain
the spin polarization observables of the photodisintegration 
of deuteron.

\subsection{Lagrangian} 

Effective Lagrangian in the pionless EFT up to NNLO
can be written as~\cite{Ando:2004mm}
\bea
{\cal L} &=& 
{\cal L}_N
+{\cal L}_s
+{\cal L}_t
+{\cal L}_{st}\,,
\eea
where ${\cal L}_N$ is a standard one-nucleon Lagrangian,
${\cal L}_{s,t}$ is a Lagrangian for two-nucleon part in 
$s$-wave spin singlet and triplet channel, respectively, and 
${\cal L}_{st}$ is a Lagrangian for spin mixing channel.
Thus, one has
\bea 
{\cal L}_N &=& N^\dagger \left[ i D_0+\frac{{\vec D}^2}{2M}
               +\frac{e}{2M}(\kappa_0+\kappa_1\tau_3)
               {\vec \sigma}\cdot {\vec B}\right] N, \\
{\cal L}_s &=& -s_a^\dagger\left[ iD_0+ \frac{{\vec D}^2}{4M}-\Delta_s
              \right] s_a -y_s \left[ s_a^\dagger N^T P_a^{(^1S_0)} N +h.c.\right], \\
{\cal L}_t &=& -t_i^\dagger \left[ i D_0 
         +\frac{{\vec D}^2}{4M}-\Delta_t\right] t_i 
         -y_t \left[ t_i^\dagger N^T P_i^{(^3S_1)} N +h.c.\right] 
         \no & &
         -\frac{C_{sd}}{\sqrt{M\rho_d}}
          \left[ \delta_{ix}\delta_{jy}-\frac{1}{3}\delta_{ij}\delta_{xy}\right] 
          \left[ t_i^\dagger (N^T {\cal O}_{xy,j} N)+ h.c.\right] 
\,,
          \\
{\cal L}_{st} &=& \frac{e L_1}{M\sqrt{r_0 \rho_d}}
           \left[ t_i^\dagger s_3 B_i +h.c.\right],          
\eea 
where $N$ is the nonrelativistic nucleon field, and 
$s_a$ and $t_i$ are the dibaryon fields in the $^1S_0$ and $^3S_1$ states, respectively.
The dibaryon fields couple with two nucleons in each partial waves 
or with other dibaryon fields. 
Spin projection operators, $P_a^{(^1S_0)}$ and $P_i^{(^3S_1)}$, for 
the $s$-wave spin singlet and triplet channels are given as 
\bea
P_a^{(^1S_0)}=\frac{1}{\sqrt{8}}\sigma_2 \tau_2\tau_a\,,
\ \ \
P_i^{(^3S_1)}=\frac{1}{\sqrt{8}}\sigma_2\sigma_i\tau_2\,,
\eea 
where $\sigma_i$ and $\tau_a$ are the Pauli matrices for spin and isospin
spaces, respectively.
The projection operators satisfy the normalization condition
\begin{eqnarray}
{\rm Tr} \left( P^\dagger_j P_k \right) = \frac{1}{2} \delta_{jk}.
\end{eqnarray}
In addition, a spin projection operator for $sd$ wave mixing channel is
given as 
\bea 
{\cal O}_{xy,j}
&=&-\frac{1}{4}(\overleftarrow{D}_x \overleftarrow{D}_y 
               P^{(^3S_1)}_j
     +P^{(^3S_1)}_j\overrightarrow{D}_x\overrightarrow{D}_y
     -\overleftarrow{D}_x P^{(^3S_1)}_j\overrightarrow{D}_y
     -\overleftarrow{D}_y P^{(^3S_1)}_j\overrightarrow{D}_x)\,,
\eea 
where
$D_\mu$ is a covariant derivative,
$D_\mu=\del_\mu+ie Q_{\rm em} A_\mu$ with a charge operator 
$Q_{\rm em}$,
the electric charge $e$, and a photon field $A_\mu$
(and $\vec{B}=\vec{\nabla}\times \vec{A}$). 
Three parameters, $M$, $\kappa_0$, and $\kappa_1$,
appear in the one-nucleon part of the Lagrangian:
$M$ is the nucleon mass, and
$\kappa_0$ and $\kappa_1$ are magnetic momenta of the nucleon for 
isosinglet and isovector channels, respectively,
$\kappa_0=0.44$ and $\kappa_1=2.35$. 
Six parameters, $\Delta_s$, $\Delta_t$, $y_s$, $y_t$, $C_{sd}$, and $L_1$,
appear in the two-nucleon part. First four parameters are fixed by 
the effective range parameters: scattering length and effective range
for each $NN$ scattering channel. 
Thus, one has~\cite{Ando:2004mm}
\bea 
\Delta_s &=& \frac{2}{Mr_0}\left(\frac{1}{a_0}-\mu\right)\,,
\quad 
\Delta_t=\frac{2}{M\rho_d}\left(\gamma-\frac{\rho_d}{2}\gamma^2-\mu\right),
\no
y_s&=&\frac{\sqrt{8\pi}}{M\sqrt{r_0}}, 
\quad   
y_t = \frac{\sqrt{8\pi}}{M\sqrt{\rho_d}},
\eea 
where $\mu$ is a dimensional parameter from the power divergence subtraction (PDS) 
regularization scheme of the loop diagrams~\cite{Kaplan:1998tg}.
\footnote{
In the calculation of loop integrals for hadronic currents in Ref.~\cite{Christlmeier:2008ye}, 
the power divergence subtraction scheme is combined with dimensional integration in spatial dimensions after using contour integration for the energy part. For more details, one may refer to Appendix in Ref.~\cite{Christlmeier:2008ye}.}
$a_0$ and $r_0$ are scattering length and effective range
in $^1S_0$ channel, 
$a_0=-23.71~\mbox{fm}$ and $r_0=2.73~\mbox{fm}$.
$\gamma$ is the deuteron binding momentum,
$\gamma= \sqrt{MB}=45.70~\mbox{MeV}$ with 
the deuteron binding energy $B$, $B=2.225\mbox{MeV}$, and
$\rho_d$ is effective range in the deuteron channel,
$\rho_d=1.764$~fm.
The remaining two parameters, $C_{sd}$ and $L_1$, are fixed by using
asymptotic ratio $\eta_{sd}=0.0254$ of $s$/$d$ wave of deuteron wavefunction
and thermal neutron capture cross section of proton, respectively.
We obtain $C_{sd}=\frac{6\sqrt{\pi}\eta_{sd}}{\sqrt{M}\gamma^2}$
and $L_1=-4.41~\mbox{fm}$.  

\begin{figure}[tbp]
\centering
\includegraphics[width=0.8\linewidth]{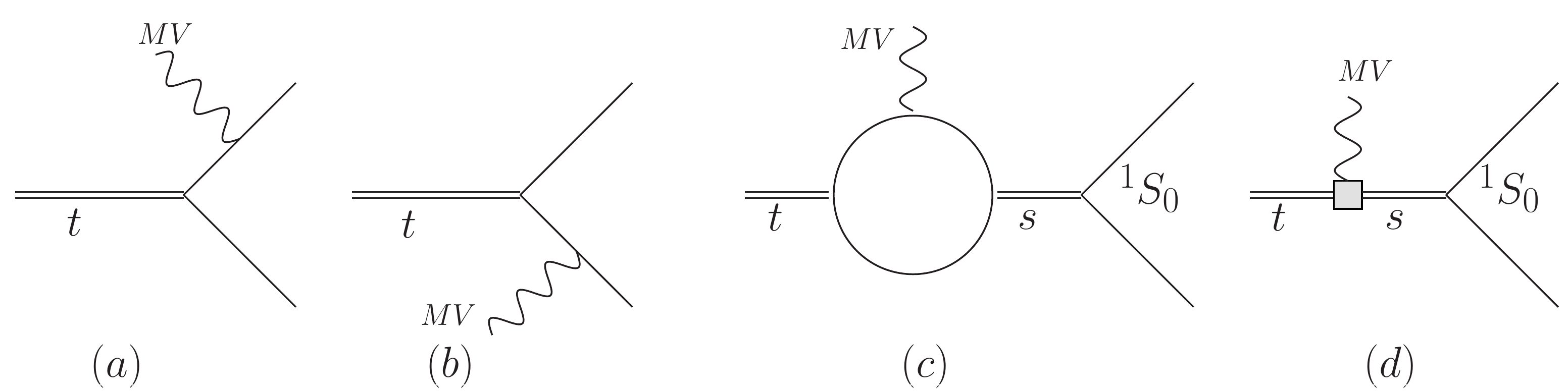}
\caption{Diagrams for LO (a-c) 
and NLO (d) of isovector magnetic currents (denoted as MV).
Double line represents dibaryon field of spin triplet (t) 
and singlet (s). 
Square of (d) represent the NLO $L_1$ term.
Only $^1S_0$ final states contribute to the magnetic currents
in (c-d) while (a-b) includes spin singlet partial waves with $L=0,1,\dots$.}
\label{fig:LO_M1}
\end{figure}

\begin{figure}[tbp]
\centering
\includegraphics[width=0.6\linewidth]{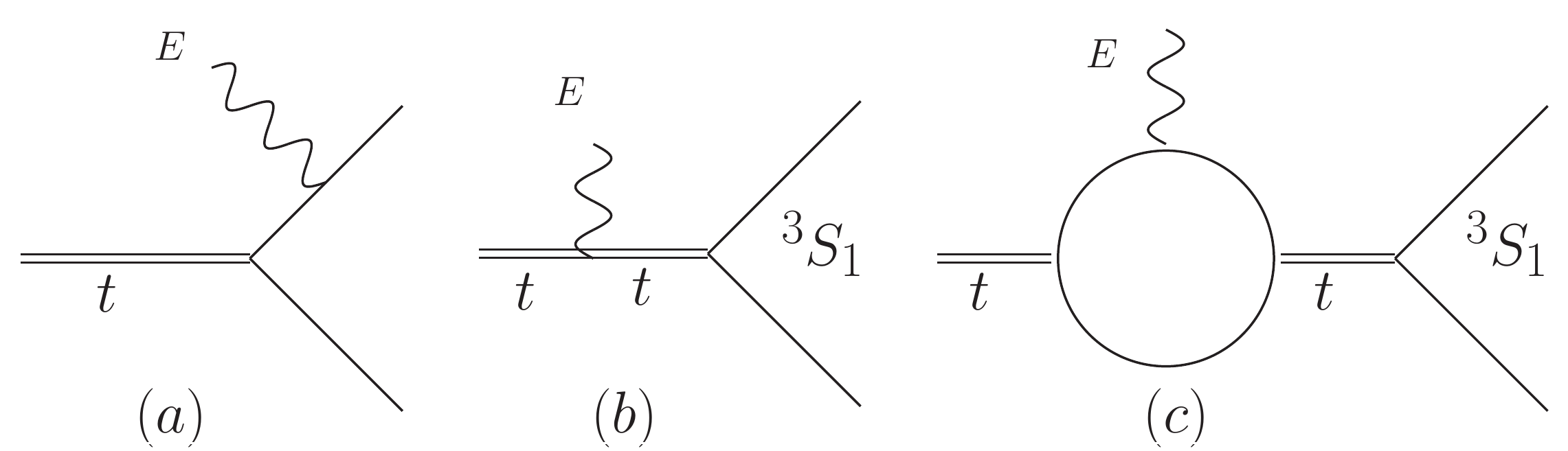}
\caption{ Diagrams for LO electric currents.
Double line represents triplet dibaryon field.
While only $^3S_1$ final state contributes in (b-c),
other partial wave contributions are included in (a)
which is dominated by the isovector E1 amplitude
to final spin triplet P-wave states.}
\label{fig:LO_E1}
\end{figure}

\begin{figure}[tbp]
\centering
\includegraphics[width=0.9\linewidth]{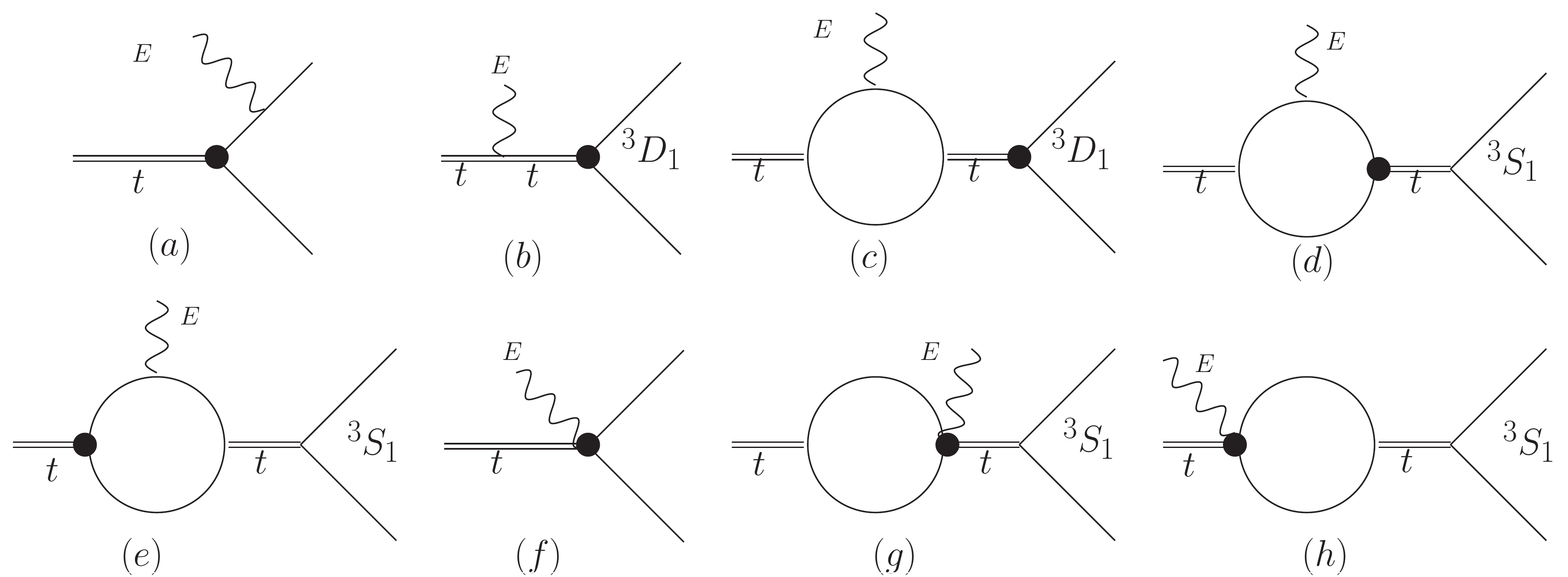}
\caption{Diagrams for NNLO electric currents (denoted as E).
Double line represents triplet dibaryon field. 
Black blob represents the $sd$ mixing $C_{sd}$ contribution. 
While only $^3S_1$ final state contributes to (d,e,g,h) and only $^3D_1$ final state
contribute to (b,c), (a,f) include other spin triplet partial waves contributions too. }
\label{fig:N2LO_E1}
\end{figure}

The amplitude ${\cal A}$ of photodisintegration, $\gamma+d\to n+p$,
can be written as
\bea 
{\cal A}= \epsilon_\mu^{(\gamma)}(q) J^{\mu}_{hadr}  
\eea 
where $\epsilon_\mu^{(\gamma)}(q)$ is a polarization vector of
the photon with four-momentum $q$, and  $J^\mu_{hadr}$ is a hadronic
current. 
Hadronic current $J_{hadr}^\mu$ is divided into electric and magnetic currents, 
$J_{hadr}^{(E)\mu}$ and $J_{hadr}^{(M)\mu}$,
which are given as 
\bea 
J^{(E)\mu}_{hadr}&=& i \frac{\sqrt{Z}}{\sqrt{8}} 
(N_p^\dagger \sigma^i\sigma_2 N_n^*)
       \epsilon_{(d)}^j J_{E, ij}^{\mu},\no 
J^{(M)k}_{hadr}&=& \frac{\sqrt{Z}}{\sqrt{8}} \epsilon^{ijk} \epsilon_{(d)}^i 
         (N_p^\dagger \sigma_2 N_n^*) J^{j}_M,         
\label{eq:hadroncurrent} 
\eea 
where $N_p$ and $N_n$ are the Pauli spinors of the proton and neutron,
and $Z$ is a wavefunction normalization factor of the deuteron,   
$Z=\frac{\gamma\rho_d}{1-\gamma\rho_d}$. 
Since the isoscalar magnetic moment is
smaller than the isovector magnetic moment
($\kappa_0/\kappa_1 \lesssim Q$), isoscalar magnetic currents are treated as
a numerically higher order in this work and thus not included in the currents.
The hadronic magnetic currents consist of contributions at leading order (LO) 
and NLO (See Fig.~\ref{fig:LO_M1}), while the electric currents have LO 
(see Fig.~\ref{fig:LO_E1}) and NNLO (see Fig. \ref{fig:N2LO_E1}) contributions 
which come from the mixing of $s$ wave and $d$ wave.
The explicit expression of the hadronic currents 
in Eq.~(\ref{eq:hadroncurrent}) 
can be found in the appendix of Ref.~\cite{Christlmeier:2008ye}.   
\footnote{Since no explicit multipole expansion is done 
and plane wave is used for final nucleons,
the currents in Eq.~\eqref{eq:hadroncurrent} contains 
various multipole amplitudes 
including dominant $M1$ and $E1$ amplitudes.}

\subsection{Cross section and observables}

An unpolarized differential cross section for the $\gamma+d\to n+p$ 
process is given 
in terms of an amplitude ${\cal A}_{m_n,m_p, \lambda,m_d}$ as
\bea
\frac{d\sigma_0}{d\Omega_p}
=\frac{\alpha}{4\pi}\frac{p_p E_p}{E_\gamma}
           \frac{1}{2\cdot 3}
 \sum_{m_n,m_p, \lambda,m_d}|{\cal A}_{m_n,m_p, \lambda,m_d}|^2\,,
\eea
where $\lambda$ denotes the polarization of photon and,
 $m_d$, $m_n$, $m_p$ are deuteron, neutron, and proton spin projections.
The unpolarized differential cross section 
in the center of mass frame with non-relativistic approximation 
is obtained by summing the spin states.
$\alpha$ is the fine structure constant,
$E_\gamma$ is the photon energy in the center of mass frame. 
The magnitude of three-momentum $p_p$ and 
the energy $E_p$ of the proton read
\bea
p_p=\frac{1}{2}\sqrt{(E_\gamma
+\sqrt{M_d^2+E_\gamma^2})^2-4 M^2},\quad E_p=\sqrt{M^2+p_p^2}\,,
\eea
where $M_d$ is the deuteron mass.
From now on, the spin indices of the amplitude will be implied 
for the amplitude ${\cal A}$ for convenience.   
Any spin dependent observables can be written in appropriate combination 
of differential cross section 
$\frac{d\sigma}{d\Omega_p}({m_n, m_p, \lambda, m_d})$. 

In this work, we will focus on the spin polarization
observables: the induced polarization $P_{y'}$ of the neutron,
and the tensor analyzing powers 
$T_{20}$ and $T_{22}$ in the deuteron photodisintegration. 
Let us choose 
the incoming photon momentum as $z$-direction $\hat{q}=\hat{z}$ which is 
also a spin quantization axis
and then introduce a second reference frame $\hat{x}'$, $\hat{y}'$, 
and $\hat{z}'$ such that the direction of the outgoing neutron momentum 
is along the $z'$ axis, $\hat{z}'=\hat{p}_n$, 
and $\hat{y}'\propto \hat{q}\times \hat{p}_n$ for convenience. 
In the form of components, $\hat{x}'$, $\hat{y}'$ and $\hat{z}'$ are 
represented as
$(\cos\theta_n\cos\phi_n,\cos\theta_n\sin\phi_n,-\sin\theta_n)$,
$(-\sin\phi_n,\cos\phi_n,0)$ and 
$(\sin\theta_n\cos\phi_n,\sin\theta_n\sin\phi_n,\cos\theta_n)$, respectively.
Thus, $P_{y'}$ is defined in terms of polarized differential cross sections,
\bea 
P_{y'}(\theta_n)&\equiv& \frac{\sigma_{+y'}(\theta_n)-\sigma_{-y'}(\theta_n)}
    {\sigma_{+y'}(\theta_n)+\sigma_{-y'}(\theta_n)}\,,
\eea  
where $\sigma_{+ y'}(\sigma_{- y'})$ 
is a differential cross section in which the spin of outgoing neutron
is parallel (anti-parallel) to the $\hat{y}'$ direction.

General form of polarized deuteron cross section with unpolarized 
photons is given as~\cite{Arenhovel:2008qy}
\bea 
\frac{d\sigma}{d\Omega}
=\frac{d\sigma_0}{d\Omega}
 \left[1+\sum_{I=1,2} P_I^d \sum_{M\geq 0} T_{IM}(\theta)\cos(M(\phi_d-\phi)-\delta_{I1}\frac{\pi}{2})
       d^{I}_{M0}(\theta_d)\right],  
\eea 
where $P^d_I$ and $T_{IM}(\theta)$ are 
orientation parameters and  analyzing powers, respectively, for $I=1,2$.
$(\theta,\phi)$ represents a direction of the outgoing proton,
the deuteron is oriented in a direction $(\theta_d,\phi_d)$,
and $d^I_{M0}$ is a rotation matrix.  
When a density matrix of the deuteron is diagonal 
in a quantization axis $\rho^d_{m'm}=p_m\delta_{m'm}$ 
where $p_m$ is a probability of finding a deuteron with a spin 
projection $m$,
orientation parameters $P_I^d$ are related to
$p_m$ as 
\bea 
P_1^d = \sqrt{\frac{3}{2}}(p_1-p_{-1})\,, 
\ \ \ 
P_2^d = \sqrt{\frac{1}{2}}(1-3p_0) \,. 
\eea   
From the rotation matrix, let us define
polarized cross sections
$d\sigma^\odot$ for 
the deuteron polarization axis 
orienting to ($\theta_d=\frac{\pi}{2},\phi_d=0$),
$d\sigma ^{\uparrow}$ to 
($\theta_d=\frac{\pi}{2},\phi_d=\frac{\pi}{2}$)
and $d\sigma ^{\downarrow}$
to ($\theta_d=\frac{\pi}{2},\phi_d=-\frac{\pi}{2}$).
Thus, we have
\bea
\frac{d\sigma^z}{d\Omega}
&\equiv&\frac{d\sigma_0}{d\Omega}
 \left[1+P_2^d T_{20}(\theta)\right], \no 
\frac{d\sigma^\odot}{d\Omega}
&\equiv&\frac{d\sigma_0}{d\Omega}
 \left[1+P_2^d T_{22}(\theta)\sqrt{\frac{3}{8}}+P_2^d T_{20}(\theta)(-\frac{1}{2})\right],\no  
\frac{d\sigma^{\uparrow,\downarrow}}{d\Omega}
&\equiv&\frac{d\sigma_0}{d\Omega}
 \left[1+P_1^d T_{11}(\theta)(\mp\frac{1}{\sqrt{2}})+P_2^d T_{22}(\theta)(-\sqrt{\frac{3}{8}})
        +P_2^d T_{20}(\theta)(-\frac{1}{2})\right].
\eea 
The tensor analyzing powers, thus, 
can be obtained from the polarized cross sections
as
\bea 
T_{11}&=&
\frac{1}{\sqrt{2} P_1^d} \frac{d\sigma ^{\uparrow-\downarrow}}{d\sigma^0},\no 
T_{20}&=&
 \frac{1}{P^d_2}\left( 2-\frac{d\sigma^\odot+\frac{1}{2}d\sigma ^{\uparrow+\downarrow}}{d\sigma^0}\right) 
 =\frac{1}{P_2^d}\left(\frac{d\sigma^z}{d\sigma_0}-1\right),
 \no 
T_{22}&=&
 \frac{\sqrt{2}}{\sqrt{3} P^d_2} 
 \frac{d\sigma^\odot-\frac{1}{2}d\sigma ^{\uparrow+\downarrow}}{d\sigma^0},
\eea 
where 
$d\sigma^{\uparrow\pm \downarrow}=d\sigma^{\uparrow}\pm d\sigma^\downarrow $.
By choosing $p_{1}=1$, 
we have $P_1^d=\sqrt{\frac{3}{2}}$, $P_2^d=1/\sqrt{2}$. 
This corresponds to the choice of $\vec{\epsilon}_{(d)}$
in Eq.~\eqref{eq:hadroncurrent} such that 
$\vec{\epsilon}_{(d)}=-\frac{1}{\sqrt{2}}(0,+i,-1)$ for $d\sigma^\odot$,
$\vec{\epsilon}_{(d)}=-\frac{1}{\sqrt{2}}(-i,0,-1)$ for $d\sigma^{\uparrow}$,
$\vec{\epsilon}_{(d)}=-\frac{1}{\sqrt{2}}(+i,0,-1)$ for $d\sigma^{\downarrow}$,
and 
$\vec{\epsilon}_{(d)}=-\frac{1}{\sqrt{2}}(1,+i,0)$ for $d\sigma^{z}$.

\section{Numerical result and Discussion}
\label{sec3}

\subsection{Total and differential cross sections}

\begin{figure}
\begin{center}
\epsfig{file=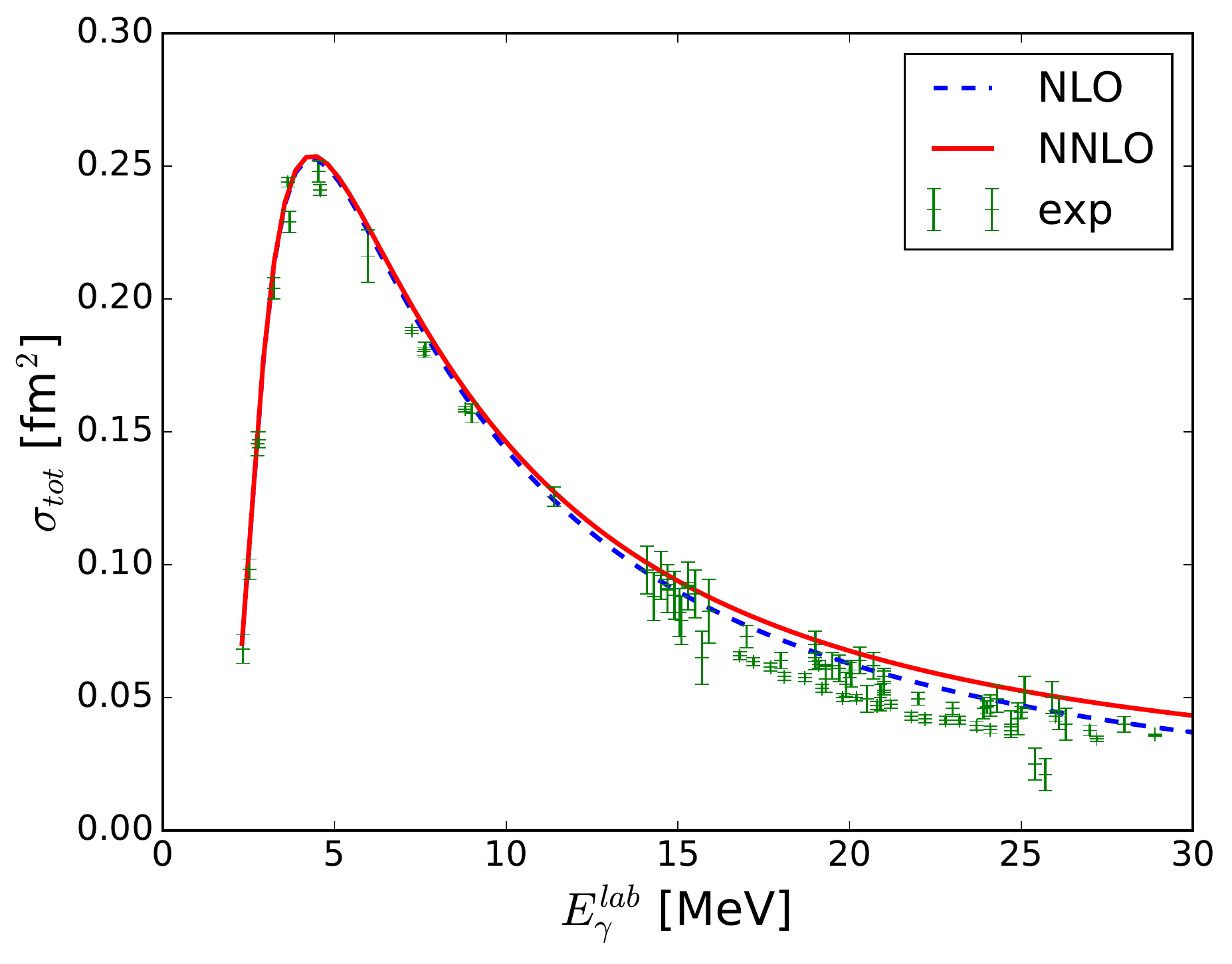, width=9cm}
\end{center}
\caption{(Color Online) Unpolarized total cross section 
$\sigma_{tot}$ of deuteron photodisintegration up to NLO and NNLO.
Experimental data~\cite{Bernabei:1988rq,Moreh:1989zz,
Bernabei:1986ai,Skopik:1974zz,Ahrens:1974ivq,Baglin:1973xoh,
Hara:2003gw,Shima:2005ix,Birenbaum:1985zz,Utsunomiya:2015bus}
are also displayed in the figure.
}
\label{fig:total}
\end{figure}

In Fig.~\ref{fig:total}, 
we plot curves of the total cross section 
calculated up to NLO 
and NNLO and include experimental 
data~\cite{Bernabei:1988rq,Moreh:1989zz,
Bernabei:1986ai,Skopik:1974zz,Ahrens:1974ivq,
Baglin:1973xoh,Hara:2003gw,Shima:2005ix,
Birenbaum:1985zz,Utsunomiya:2015bus} as well.
With the parameters fixed to low-energy data, 
the curves up to NLO and NNLO give the results consistent with the data
to $E_\gamma^{lab} = 30$~MeV. 
The correction from NNLO is about 10\% of the contribution up to NLO 
at $E_\gamma^{lab} \sim 20$~MeV, and about 20\% 
at $E_\gamma^{lab} \sim 30$~MeV. 
Although the total cross section turns out to be 
in good agreement with the experimental data 
up to relatively high energy,  in principle, 
the pionless theory must be applied 
to the photon energies below 10~MeV which corresponds
to nucleon momentum close to the pion mass. 
Thus, in this work, we limit
the photon energy to the range $E^{lab}_\gamma \leq 20$~MeV
in the study of the spin observables.

\begin{figure}
\begin{center}
\epsfig{file=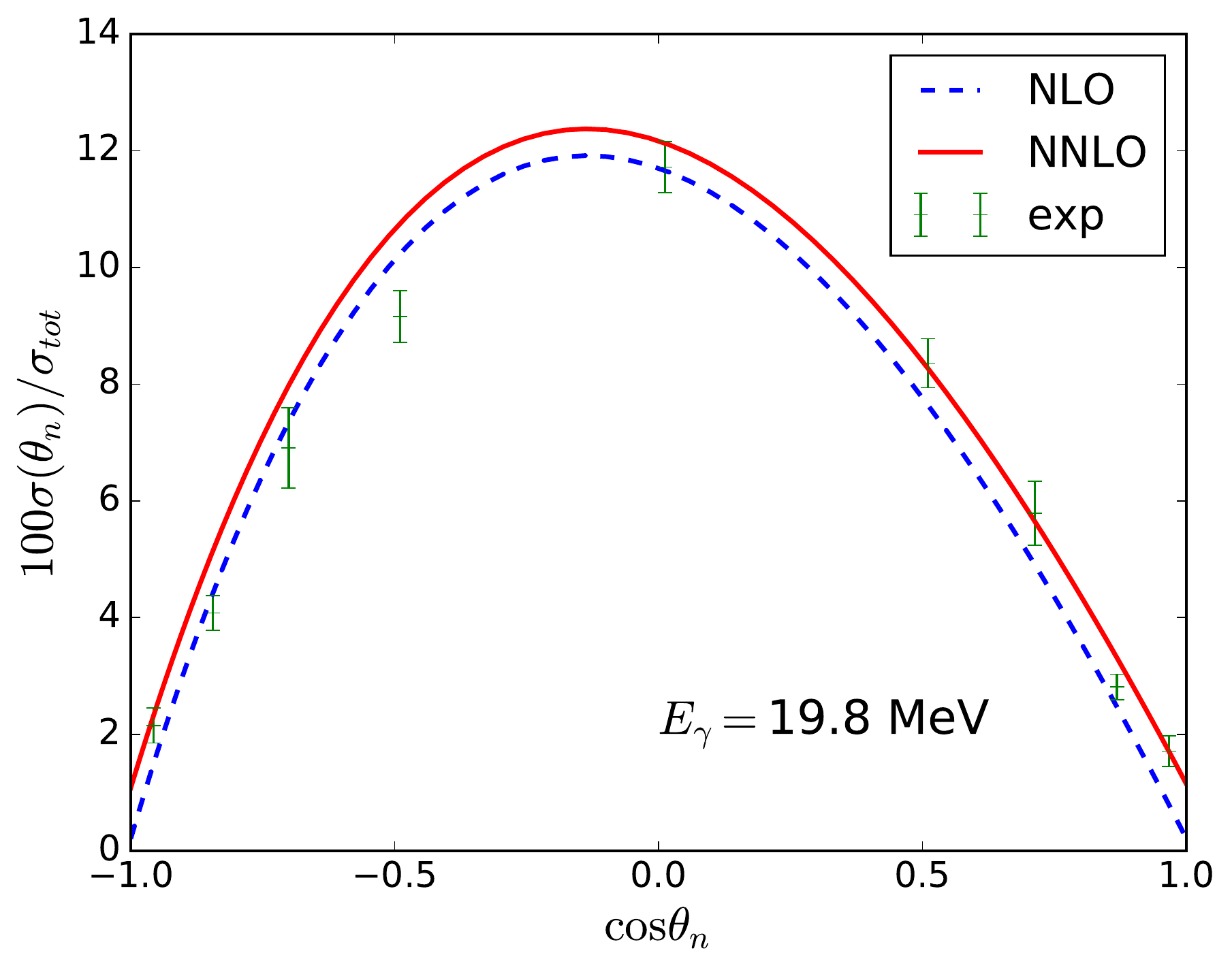, width=8cm}
\end{center}
\caption{(Color Online)  
Ratio of unpolarized differential cross section and total cross section
for the photon energy 19.8~MeV up to NLO (blue dashed) and NNLO (red solid). 
Experimental data~\cite{DePascale:1985np} are also displayed. 
}
\label{fig:diff}
\end{figure}

In Fig.~\ref{fig:diff}, curves of the unpolarized differential cross 
section divided by the total cross section 
at $E_\gamma = 19.8$~MeV
are plotted by using our results up to NLO and NNLO,
and the experimental data are also included~\cite{DePascale:1985np}. 
Similar to the total cross section, 
the difference between NLO and NNLO is negligible 
up to 10~MeV, and the NNLO correction converges well 
even at $E_\gamma = 19.8$~MeV.

\subsection{Induced neutron polarization $P_{y'}$}

\begin{figure}
\begin{center}
\epsfig{file=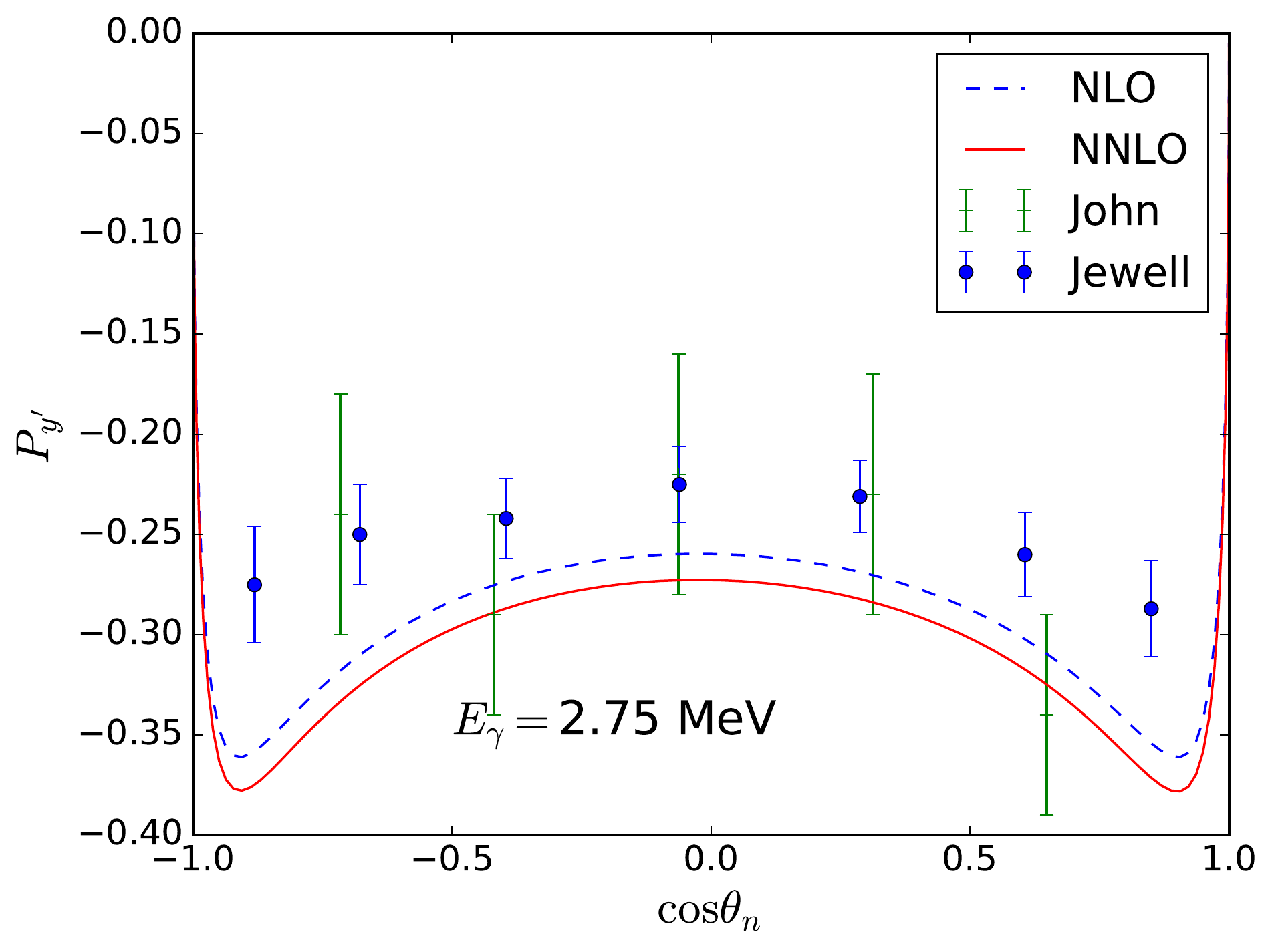, width=8cm}
\epsfig{file=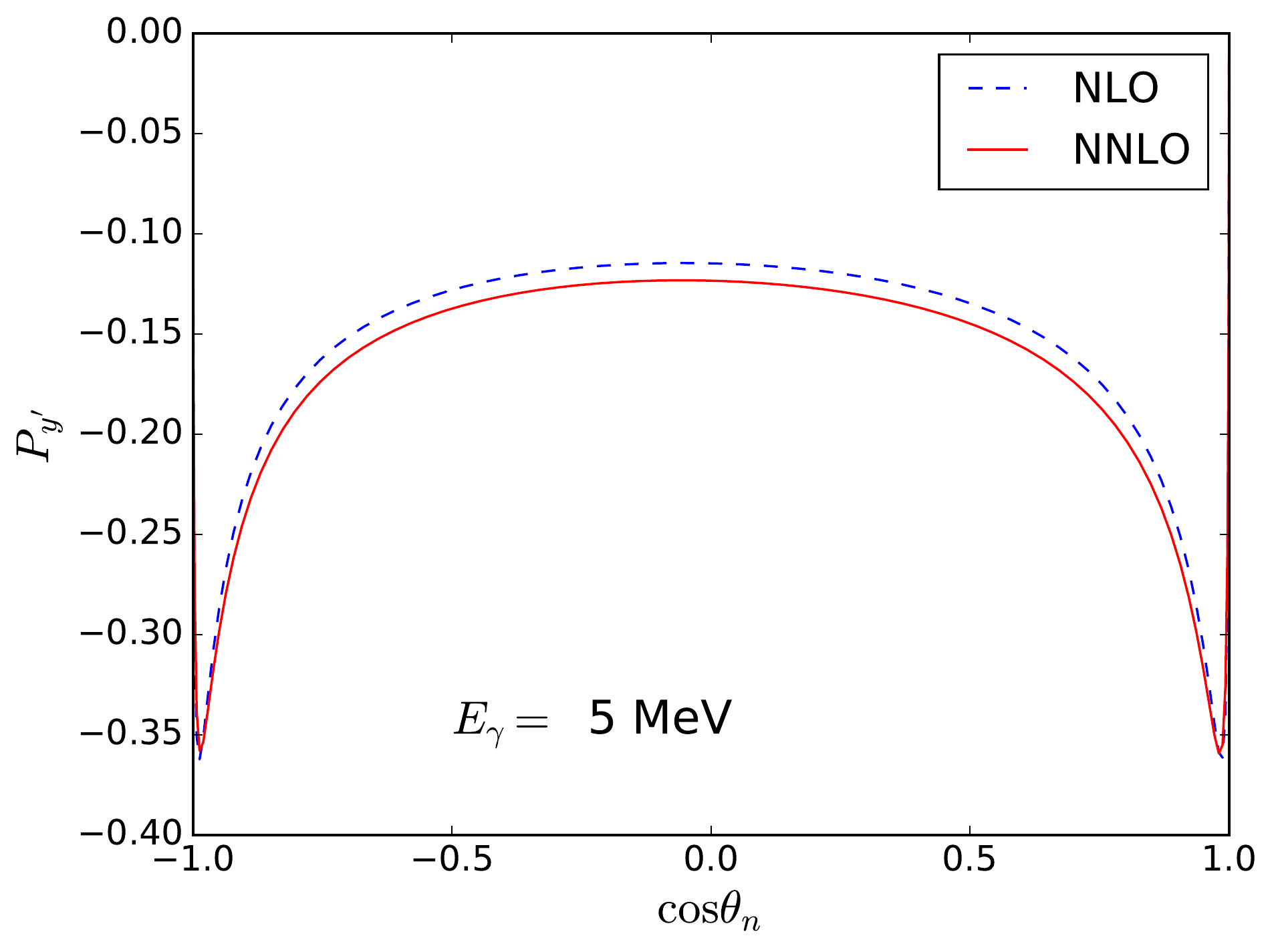, width=8cm}
\epsfig{file=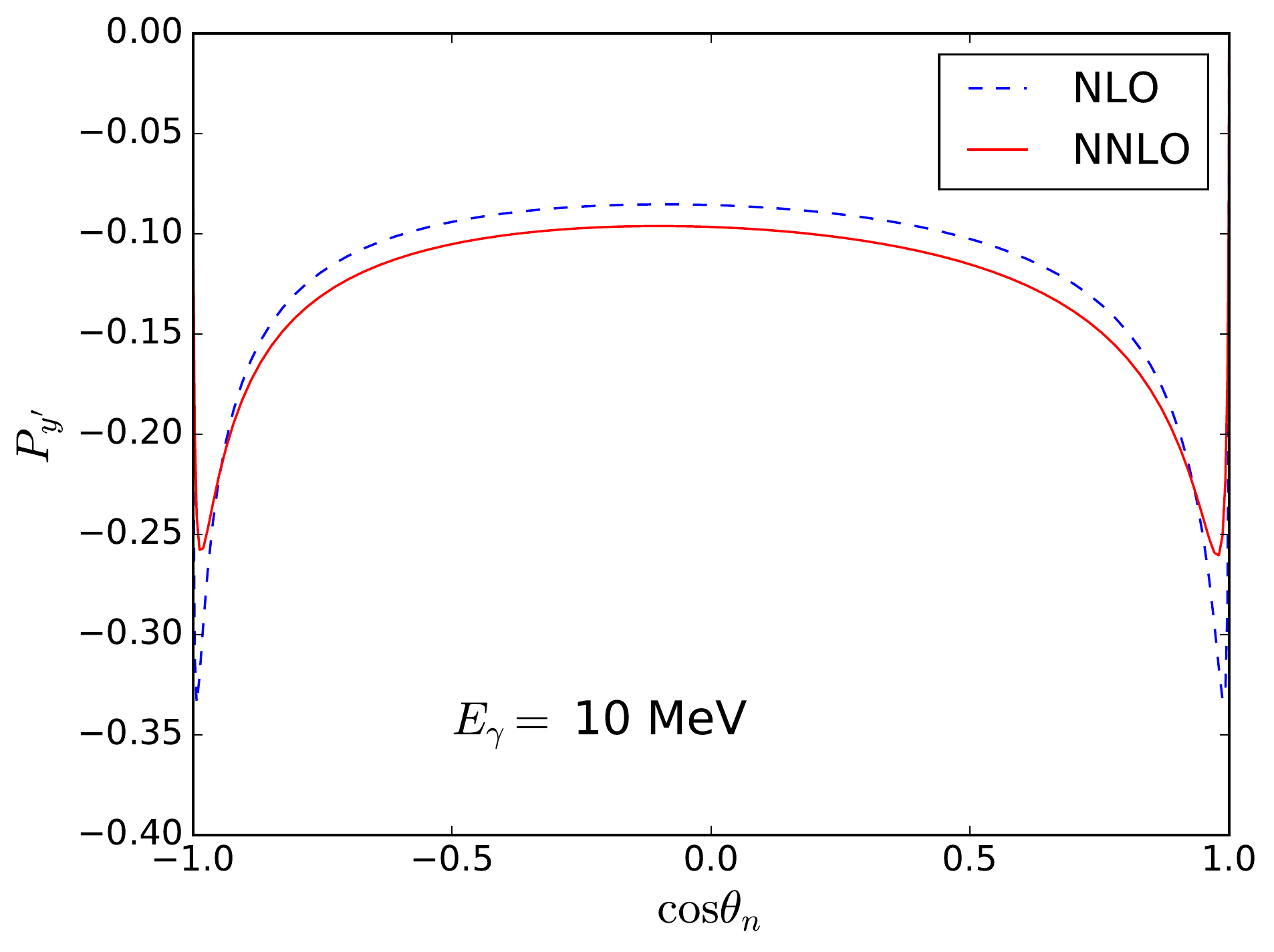, width=8cm}
\epsfig{file=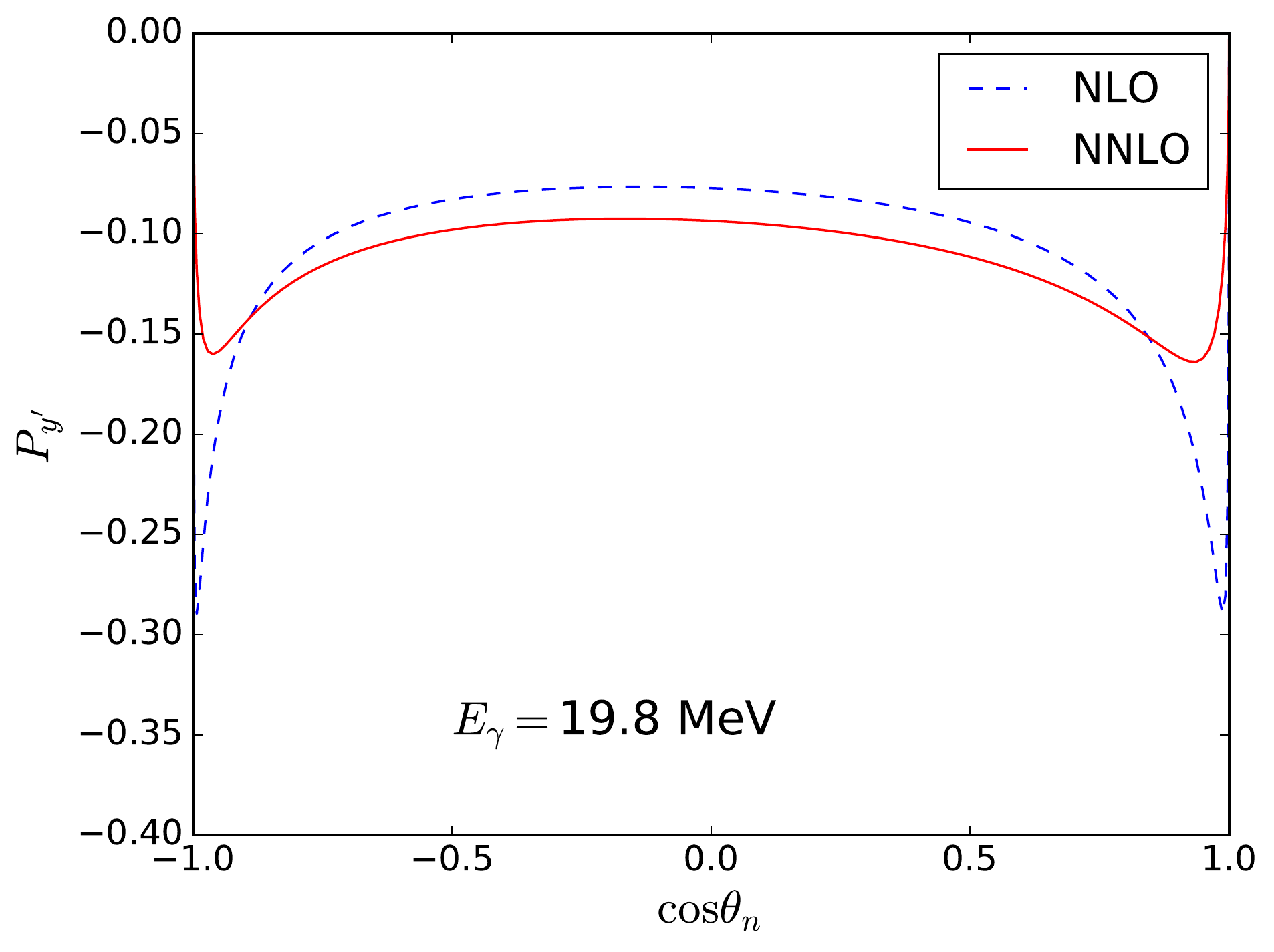, width=8cm}
\end{center}
\caption{(Color Online) $P_{y'}$ for the photon energies 2.75 MeV (top left), 
5 MeV (top right), 10 MeV (bottom left), and 19.8 MeV (bottom right) up to
NLO (blue dashed) and NNLO (red solid).
Experimental data labeled by ``John"~\cite{PhysRev.124.830} and 
``Jewell"~\cite{PhysRev.139.B71} are included in the figure at 
$E_\gamma = 2.75$~MeV.
}
\label{fig:pyp:ang}
\end{figure}

In Fig.~\ref{fig:pyp:ang},
our results of $P_{y'}$ up to NLO and NNLO 
at $E_\gamma = 2.75$, 5, 10, 19.8~MeV are plotted as functions
of $\cos\theta_n$, where $\theta_n$ is the angle for the outgoing neutron
in the center of mass frame. 
Experimental data labeled by 
``John''~\cite{PhysRev.124.830} and ``Jewell''~\cite{PhysRev.139.B71}
at $E_\gamma=2.75$~MeV 
are also included in the figure. 
We find that the NNLO corrections converge well 
at all the photon energies and are small compared 
to those up to NLO. 
Discrepancy between theory and experiment
of $P_{y'}$ at $E_\gamma=2.75$~MeV, which has been reported 
in our previous work in the pionless EFT up to NLO~\cite{Ando:2011nv},
cannot be resolved by including the NNLO corrections. 

\begin{figure}
\begin{center}
\epsfig{file=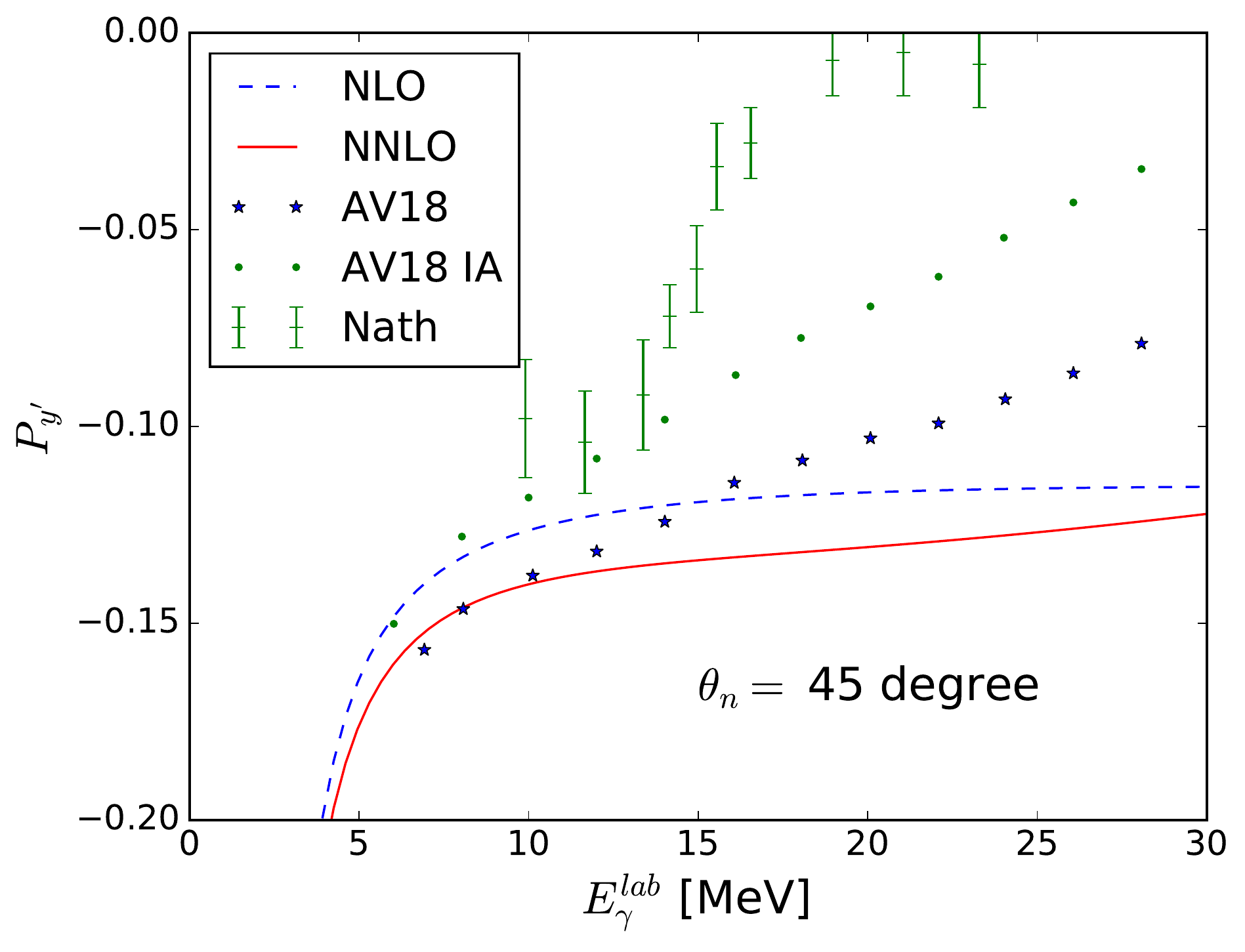, width=8cm}
\epsfig{file=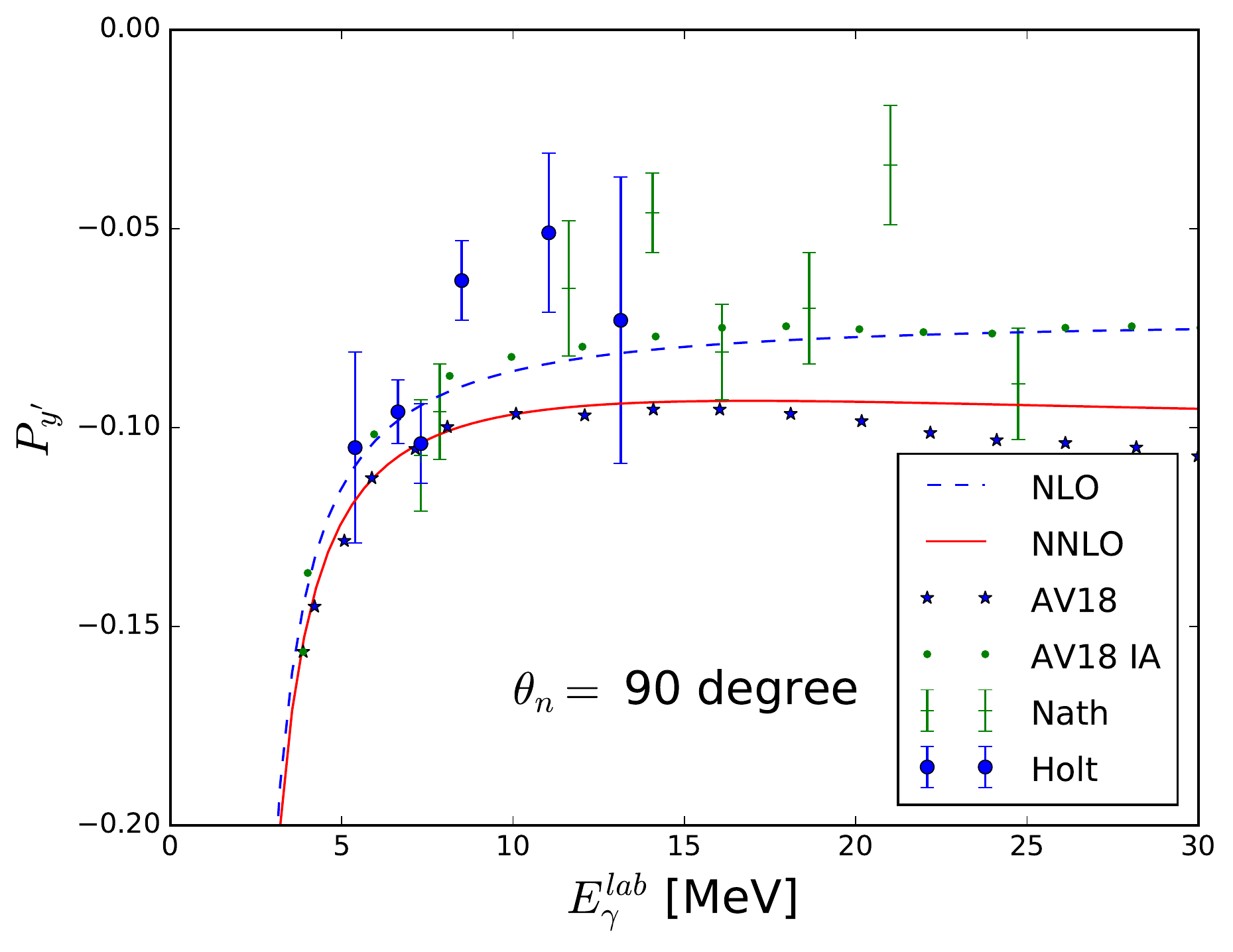, width=8cm}
\epsfig{file=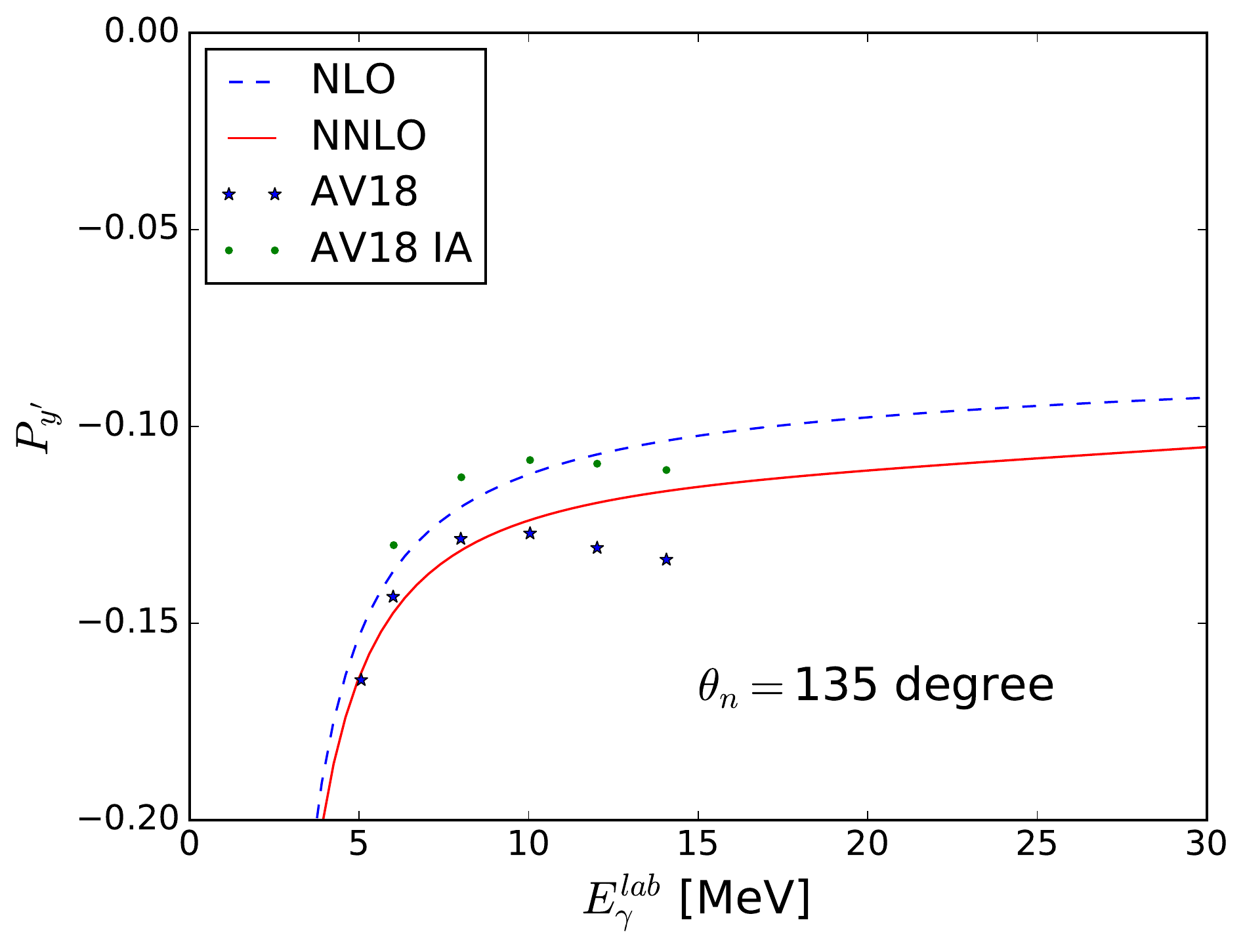, width=8cm}
\end{center}
\caption{(Color Online) $P_{y'}$ at three different  
angles with increasing photon energy  up to
NLO (blue dashed) and NNLO (red solid).
Top left is for $\theta_n= 45^\circ$, 
top right is for $\theta_n= 90^\circ$ 
and the bottom is for $\theta_n= 135^\circ$. 
Experimental results from Nath \cite{Nath:1972akq} 
and Holt \cite{Holt:1983zza} are also displayed
in the figures at $\theta_n=45^\circ$ and 90$^\circ$. 
}
\label{fig:pyp:en}
\end{figure}

Figure \ref{fig:pyp:en} shows the results of $P_{y'}$ in three different  
angles with increasing photon energy. 
The pionless EFT results up to NLO
and NNLO agree well with phenomenological potential model calculation
of Av18 in impulse approximation and Av18 calculation
with exchange currents~\cite{Schiavilla:2005zt}
at low energies, respectively. 
The agreement implies that model-independent calculation of
pionless EFT supports the Av18 results rather than the measurement by 
Nath~\cite{Nath:1972akq}.
The results of Av18 and pionless EFT converge up to $E_\gamma \sim 10$~MeV,
and start to deviate for $\theta_n = 45^\circ$ and $135^\circ$ 
while $\theta_n = 90^\circ$ results
show good agreement even at $E_\gamma \geq 10$~MeV.
This could be related with higher order corrections 
which are neglected in the present work.

\subsection{Tensor analyzing power $T_{20}$ and $T_{22}$}

By computing differential cross sections,  
$d\sigma^\odot$, $d\sigma ^{\uparrow}$, 
and $d\sigma ^{\downarrow}$, 
we can obtain the tensor analyzing powers $T_{11}$, $T_{20}$
and $T_{22}$. 
However, because $d\sigma ^{\uparrow}\simeq d\sigma ^{\downarrow}$
up to NNLO,
we obtain $T_{11}\simeq 0$. 
We would need to include higher order corrections 
to have a sizable contribution to $T_{11}$.  
Also, $T_{22}$ is dominated by the NNLO contribution since 
$T_{22}\simeq 0$ at NLO, which 
implies a contribution to $T_{22}$ mostly comes 
from the $sd$ mixing effects.

\begin{figure}
\begin{center}
\epsfig{file=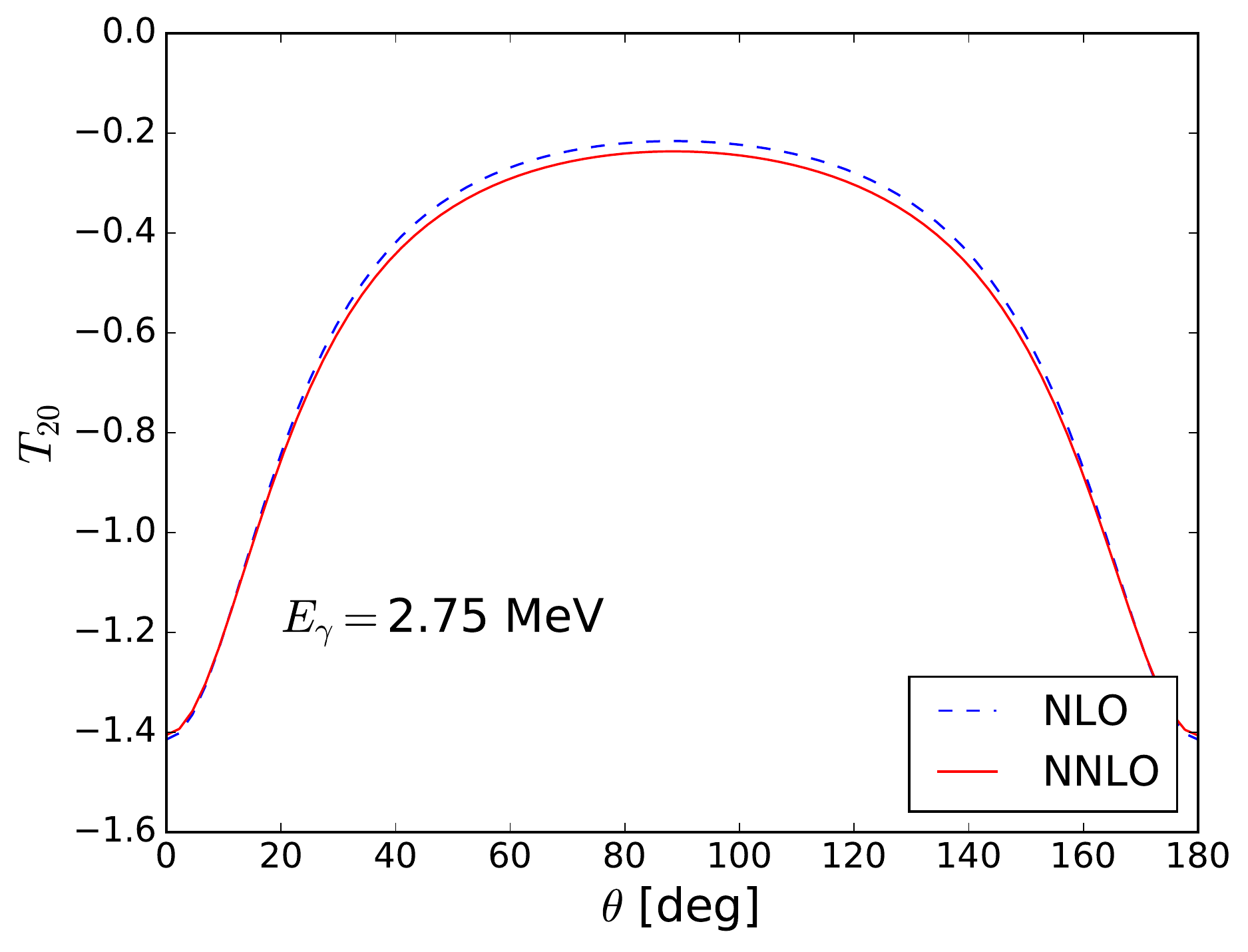, width=8cm}
\epsfig{file=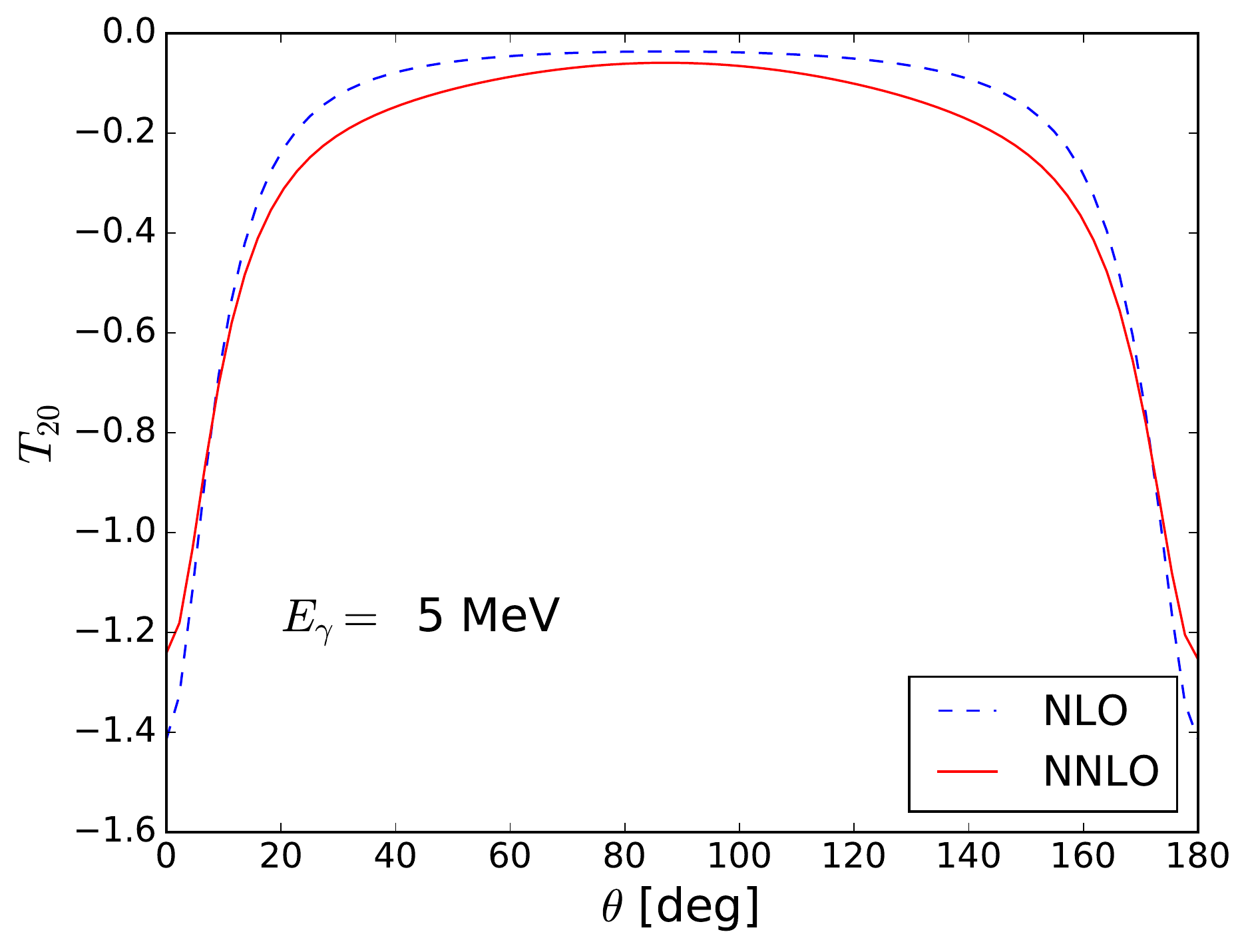, width=8cm}
\epsfig{file=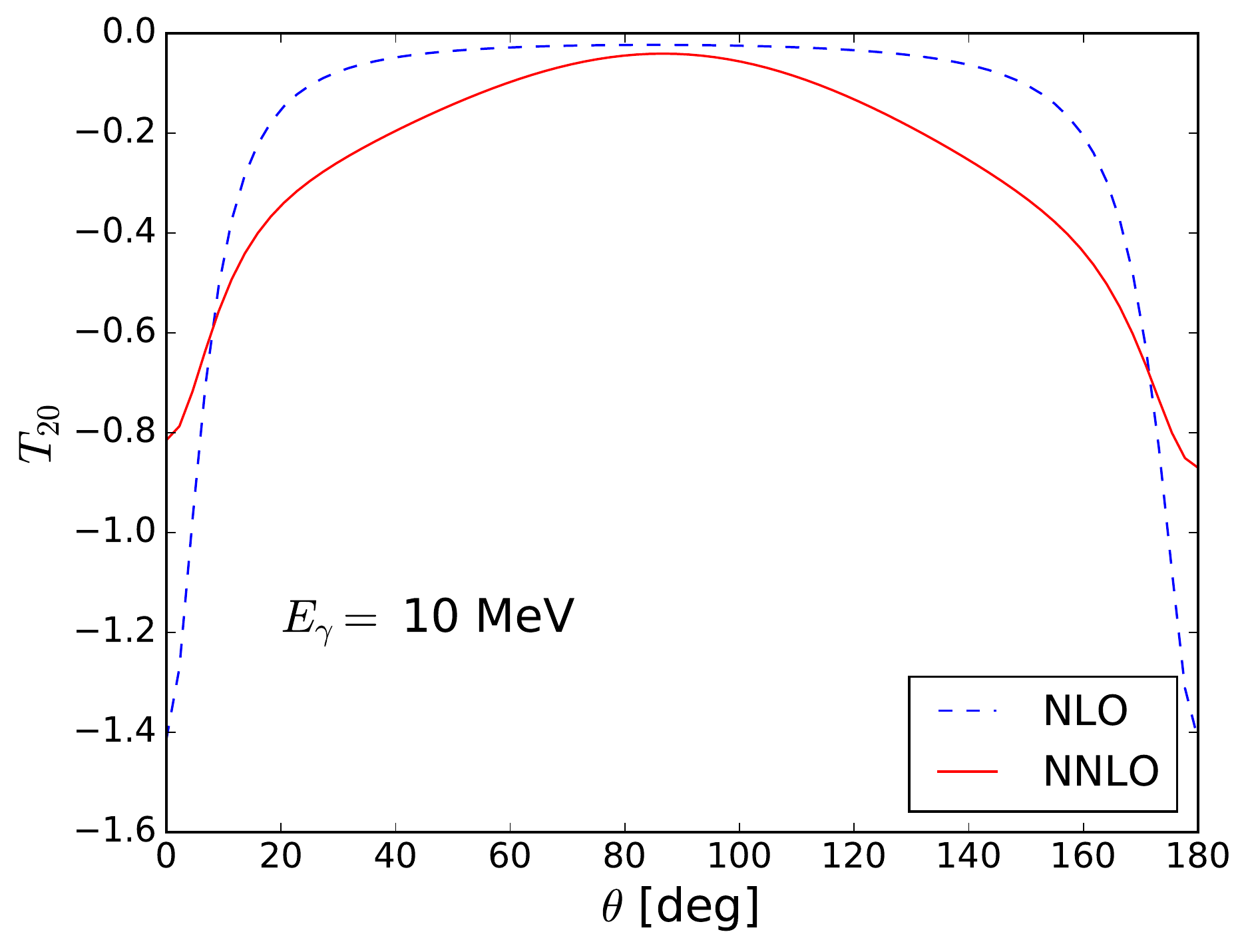, width=8cm}
\epsfig{file=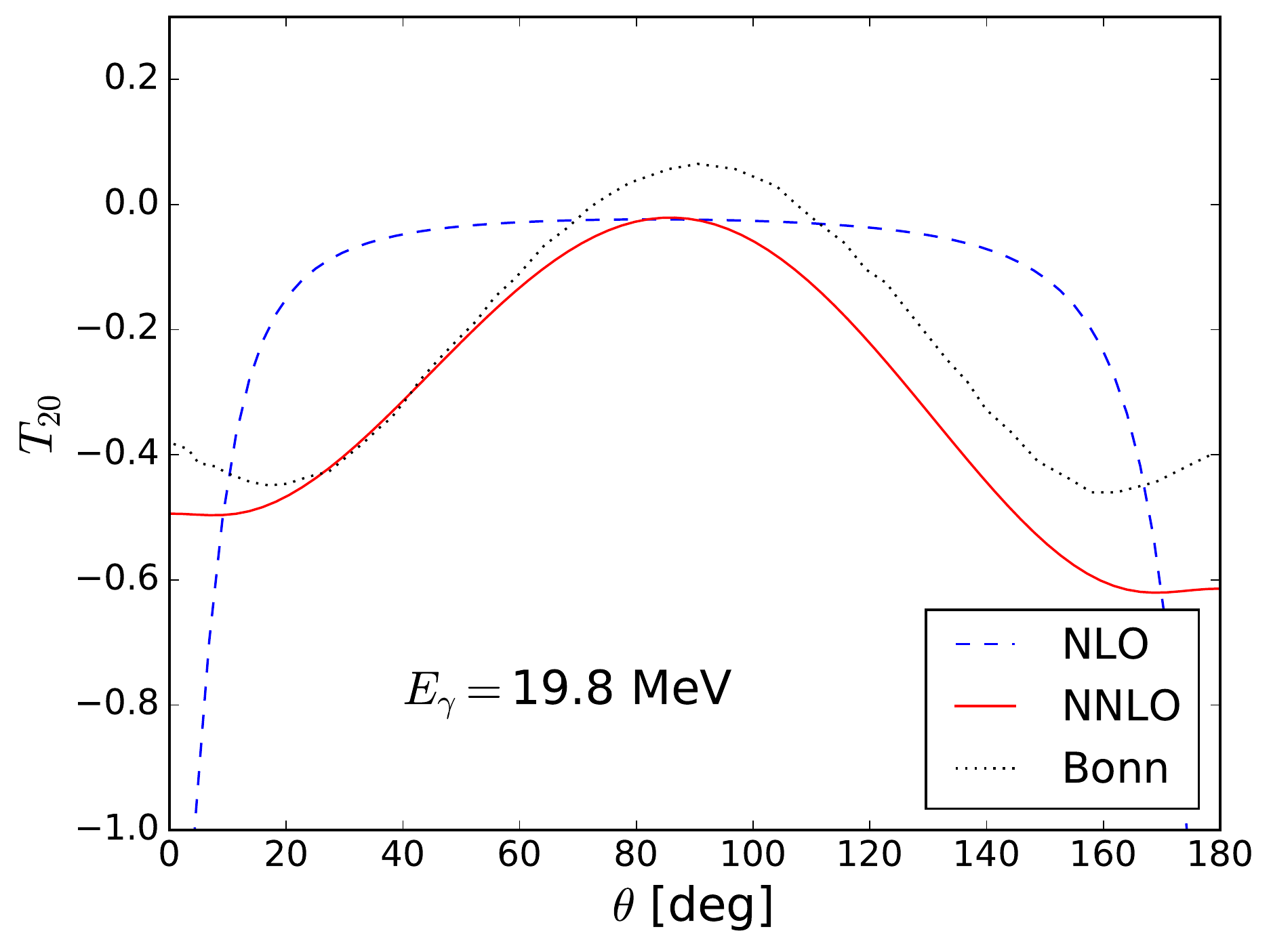, width=8cm}
\end{center}
\caption{(Color Online) $T_{20}$ for the photon energies 2.75 MeV (top left), 
5 MeV (top right), 10 MeV (bottom left), and 19.8 MeV (bottom right)
up to NLO (blue dashed) and NNLO (red solid).
Bonn potential result is from Ref.~\cite{Arenhovel:1990yg} (black dotted).}
\label{fig:T20}
\end{figure}

In Figure \ref{fig:T20}, we summarize our results of $T_{20}$ at NLO and NNLO.
There are interesting behaviors 
absent or appearing weakly in other observables.
First of all, 
in the observables considered so far, the results of NNLO is almost 
identical to those of NLO  up to $E_\gamma=10$~MeV, 
but for $T_{20}$ non-negligible 
difference appears already at $E_\gamma =5$~MeV.
As the energy increases, the NLO result becomes the shape of flat 
and wide plateau.
The value of $T_{20}$ in the flat region is close to zero and 
independent of the energy.
Similarly the values at $\theta=0^\circ$ and $180^\circ$ remain 
the same regardless of the energy.

The result at $E_\gamma = 19.8$~MeV 
is similar to the results with those obtained from
the Bonn potential model calculations, lying between ``N+MEC+IC'' and 
``N+MEC+IC+RC'' in Fig.~7.4.13 in \cite{Arenhovel:1990yg}.
The effect of NNLO can be distinguished from that of NLO even 
at $E_\gamma = 10$~MeV, 
which is the energy of interest in the HIGS proposal~\cite{higs}.
Therefore measurement of $T_{20}$ will provide a unique opportunity 
to test the role of higher orders in the pionless theory. 
At the same time, it might help understand the origin of discrepancy 
between experiment and theory in the polarization observables 
in the few-body systems.

\begin{figure}
\begin{center}
\epsfig{file=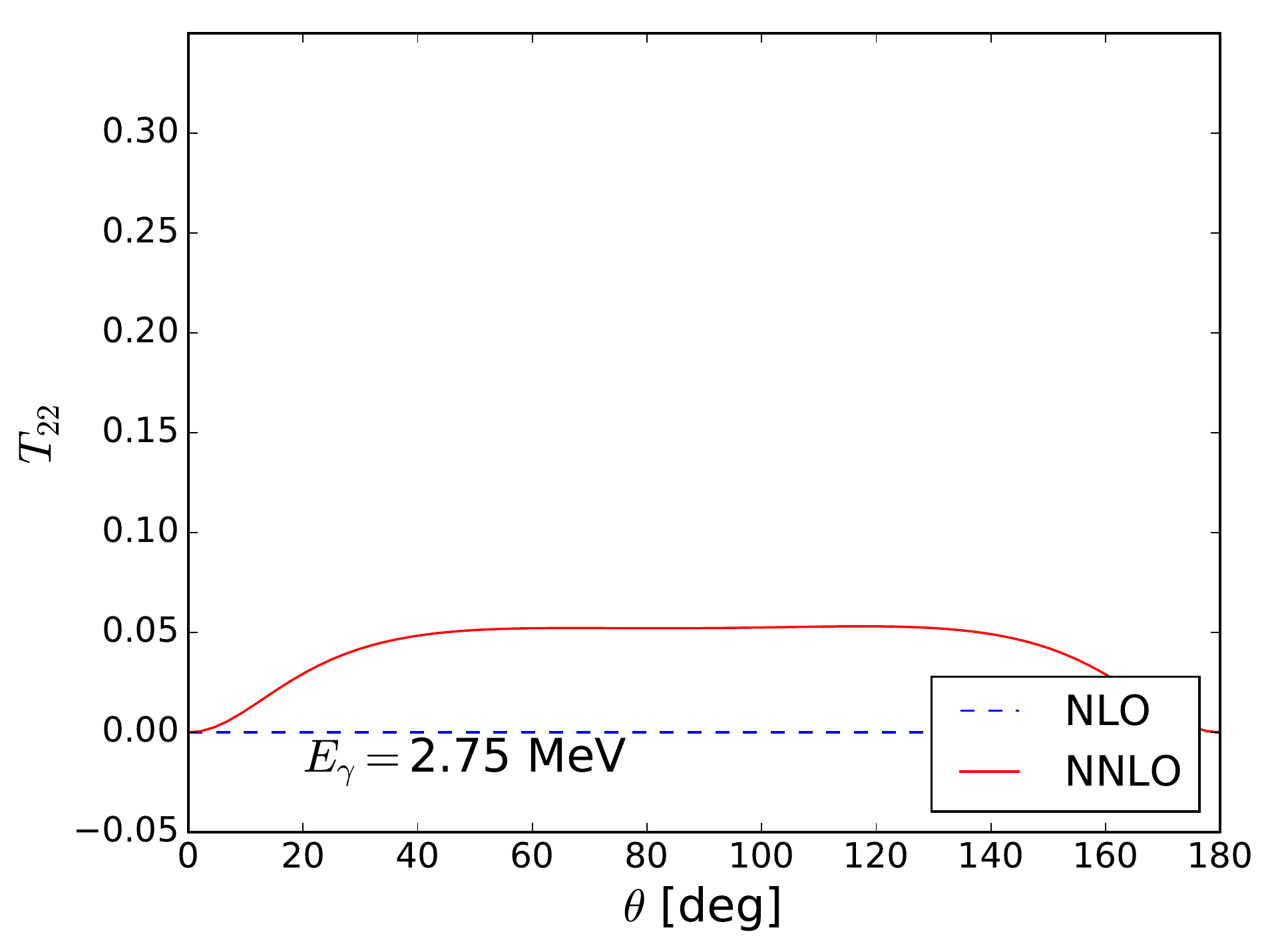, width=8cm}
\epsfig{file=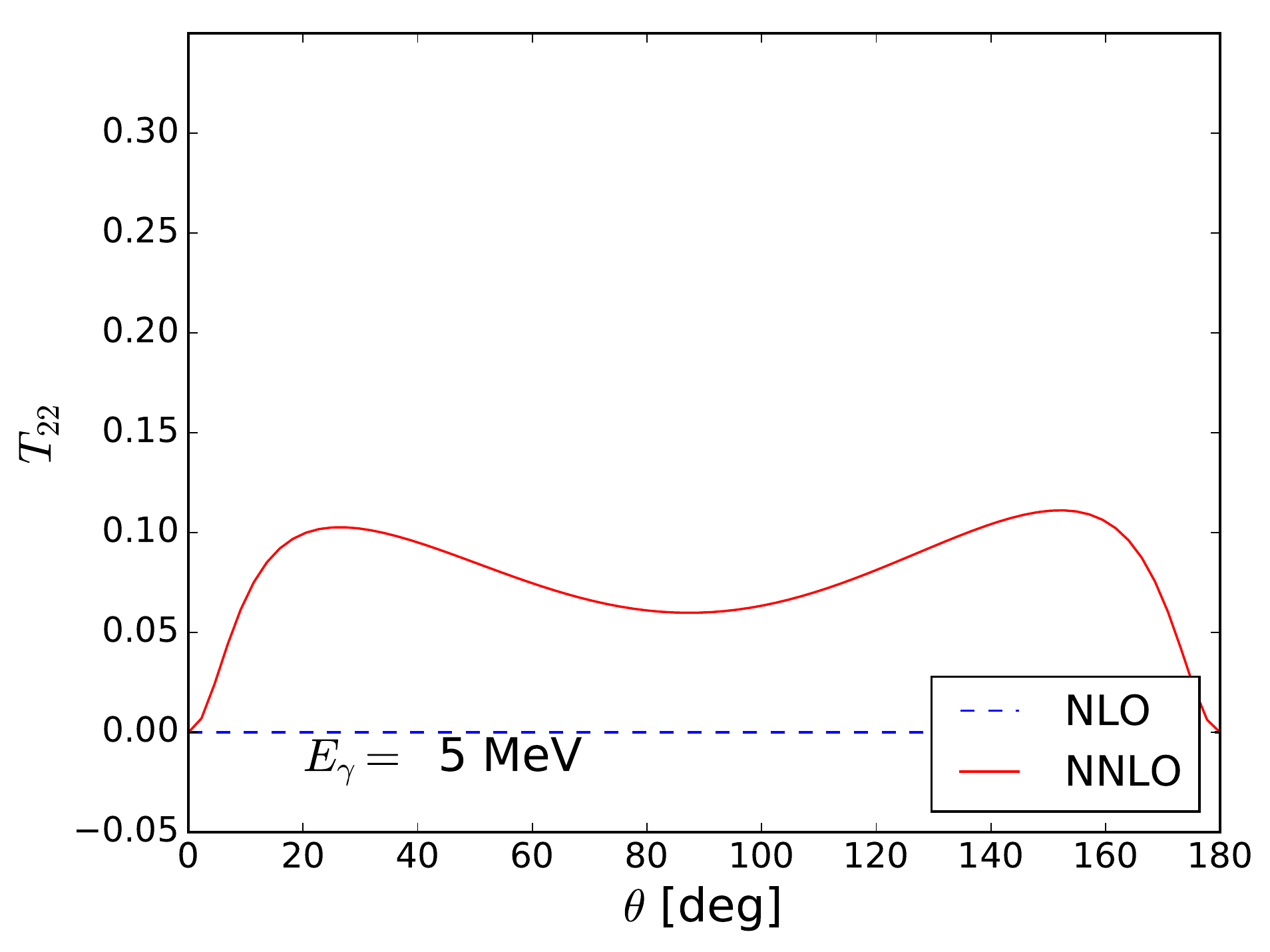, width=8cm}
\epsfig{file=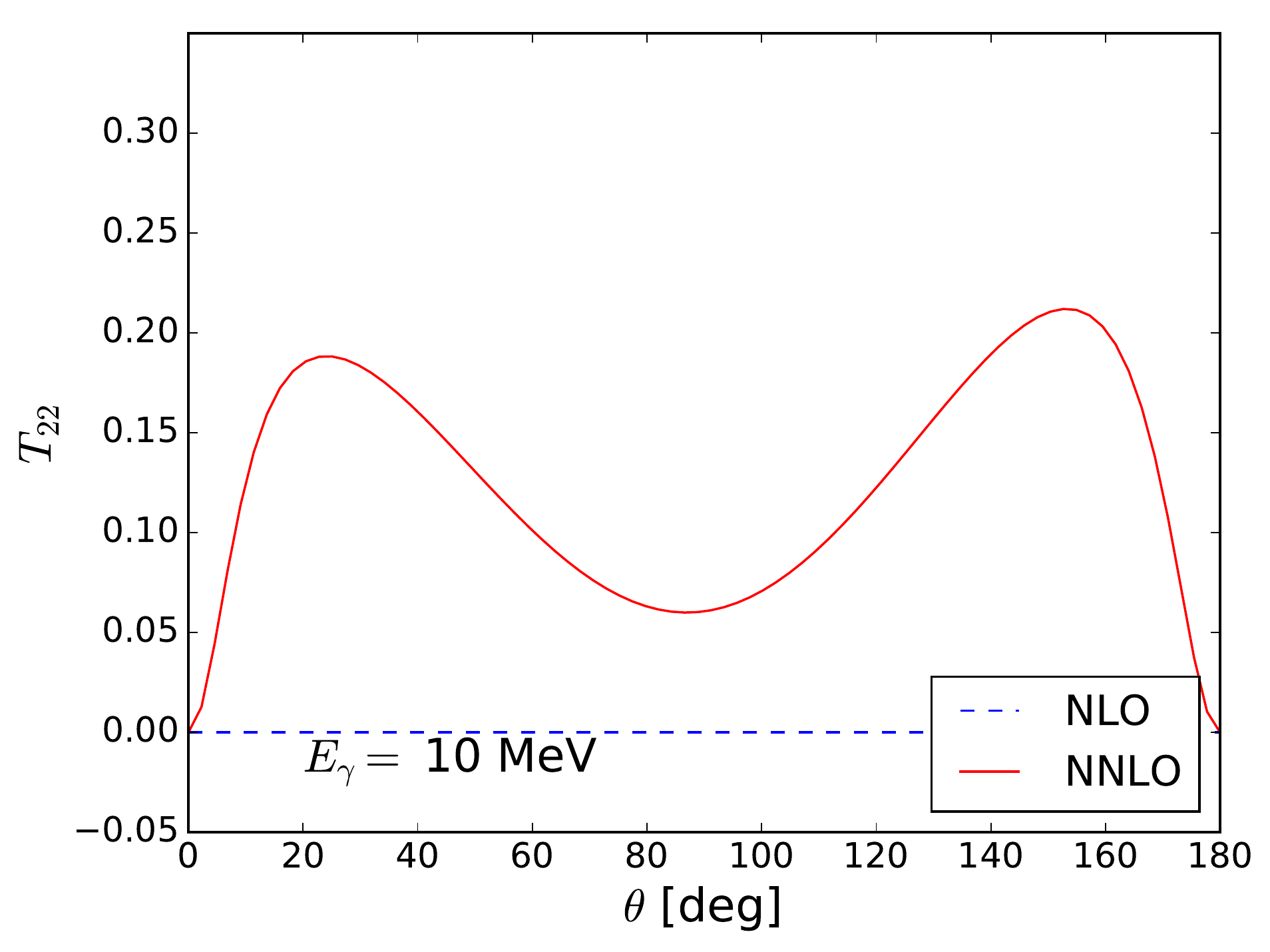, width=8cm}
\epsfig{file=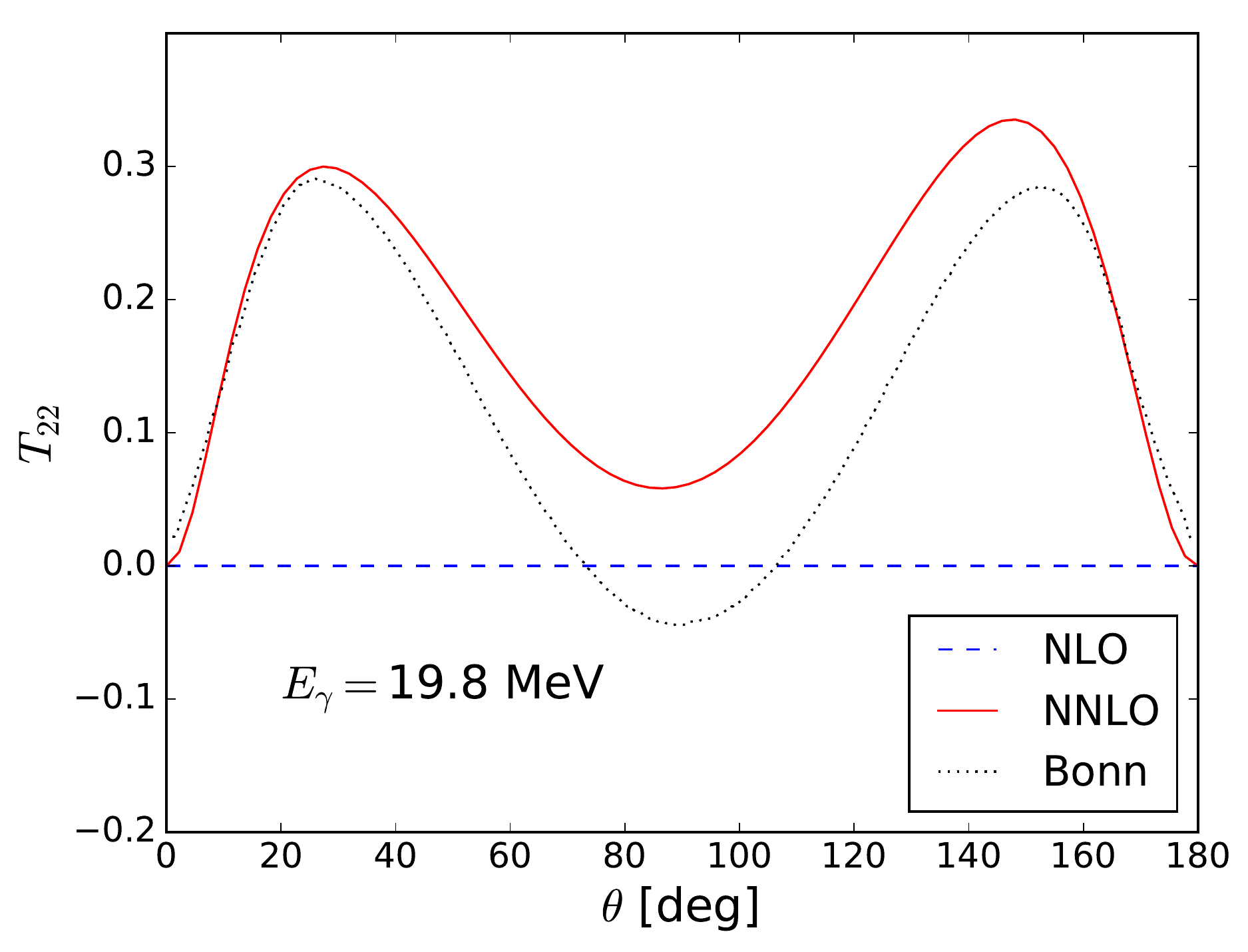, width=8cm}
\end{center}
\caption{(Color Online) $T_{22}$ for the photon energies 2.75 MeV (top left), 
5 MeV (top right), 10 MeV (bottom left), and 19.8 MeV (bottom right)  up to
NLO (blue dashed) and NNLO (red solid).
Bonn potential result is from Ref.~\cite{Arenhovel:1990yg} (black dotted).}
\label{fig:T22}
\end{figure}

In Fig.~\ref{fig:T22}, curves of $T_{22}$ up to NLO and NNLO at
$E_\gamma = 2.75$, 5, 10, 19.8~MeV are plotted as functions of 
$\theta$. As mentioned before, there is no contribution from 
the NLO corrections, and the NNLO corrections, mainly the $sd$ wave 
mixing term, are a leading contribution to $T_{22}$ at the low energies. 
Our result at $E_\gamma =19.8$~MeV is compared to a result of 
the Bonn potential model~\cite{Arenhovel:1990yg}.
One can see quantitative agreement between the NNLO result 
and the Bonn potential one.

\section{Summary}
\label{sec4}

Motivated by the proposal of measurement 
of the tensor analyzing power $T_{20}$ at HIGS facility, 
we studied the photodisintegration of the deuteron
at low energies.
Pionless EFT with dibaryon fields is used as the tool for calculation,
and corrections up to NNLO are included.
Various observables such as the total and differential cross sections, 
and spin-dependent observables are investigated.

For the quantities that have non-vanishing contributions from LO 
such as the cross sections
and $P_{y'}$, NNLO terms give perturbative corrections to the NLO results, 
thus the theory shows good convergent behavior.
For $P_{y'}$, including NNLO contributions, 
our result becomes closer to a sophisticated
calculation with Av18 model.
The discrepancy between measurement and NLO result remains 
unsolved even if we include the NNLO corrections.

For $T_{20}$, NNLO gives negligible change to the NLO result 
at $E_\gamma = 2.75$~MeV,
but the correction becomes more significant as the energy increases.
No data are available below 19.8~MeV, and we can make comparison to the 
result with the Bonn potential model at $E_\gamma = 19.8$~MeV.
The agreement to the Bonn model result depends on the angle, but as a whole 
the NNLO result agrees well with that of the Bonn model quantitatively.

For $T_{22}$, contributions up to NLO are null, 
and non-vanishing values appear at NNLO.
NNLO result agrees to the Bonn model result fairly well.
The agreement in the tensor analyzing power proves that 
increase of the order in both wavefunctions
and operators for the external probe in the pionless EFT 
can give results as accurate as those of the most elaborate calculation 
with modern phenomenological potential models.
Since our work includes electric and magnetic hadronic currents up to NNLO,
it would be interesting to check the effects of other multipole operators.

\section*{Acknowledgments}

We are grateful to Pil-Neyo Seo for stimulated discussion and providing us 
the information of the measurement of $T_{20}$. 
YHS and SIA thank the Institute for Nuclear Theory 
at the University of Washington for its hospitality 
during the Program INT-16-1 ``Nuclear Physics from Lattice QCD", 
where the present work was started.
Work of YHS was supported in part by
the Rare Isotope Science Project of the
Institute for Basic Science funded by Ministry of Science,
ICT and Future Planning and
National Research Foundation of Korea (2013M7A1A1075764).
That of SIA was supported by the Basic Science Research Program 
through the National Research Foundation of Korea funded 
by the Ministry of Education of Korea (Grant No.~NRF-2016R1D1A1B03930122)
and (Grant No.~NRF-2016K1A3A7A09005580). 

\bibliography{dgnp}


\end{document}